\documentclass[epjc3,twocolumn,a4paper,fleqn]{svjour3} 
\usepackage[fleqn]{amsmath}
\usepackage{graphicx}
\usepackage{epstopdf}
\usepackage{fix-cm}
\usepackage{txfonts}
\usepackage{lipsum}
\usepackage{mathtools}
\usepackage{cuted}
\usepackage{multirow}
\usepackage{comment}
\usepackage{empheq}
\usepackage[titletoc,title]{appendix}

\usepackage{listings}
\usepackage{lineno}
\usepackage{mdwlist}
\usepackage{pennames}
\usepackage[british]{babel}
\usepackage{color}
\usepackage{subfigure}
\usepackage{siunitx}
\usepackage{import}
\usepackage{amssymb}
\usepackage{epsfig}
\usepackage{cite}
\usepackage{hyperref}
\hypersetup{
colorlinks=true,
linkcolor=blue,
citecolor=blue,
urlcolor=blue,
}

\DeclareSIUnit \clight  {\textit{c}}
\RequirePackage[normalem]{ulem} 
\RequirePackage{color}\definecolor{RED}{rgb}{1,0,0}\definecolor{BLUE}{rgb}{0,0,1} 
\RequirePackage[normalem]{ulem} 
\RequirePackage{color}\definecolor{RED}{rgb}{1,0,0}\definecolor{BLUE}{rgb}{0,0,1} 

\DeclareGraphicsExtensions{.eps,.pdf,.jpeg,.jpg,.png}

\def\vector#1{\mbox{\boldmath $#1$}}

\newcommand{\bea}{\begin{eqnarray}}
\newcommand{\eea}{\end{eqnarray}}
\newcommand{\be}{\begin{equation}}
\newcommand{\ee}{\end{equation}}
\newcommand{\fref}[1]{Fig.~\ref{#1}}
\newcommand{\Fref}[1]{Figure~\ref{#1}}

\newcommand{\tref}[1]{Table~\ref{#1}}

\newcommand*{\muonp}          {\ifmmode\mathrm{\muup^+}\else$\mathrm{\muup^+}$\fi}
\newcommand*{\muon}           {\ifmmode\mathrm{\muup}\else$\mathrm{\muup}$\fi}
\newcommand*{\tauon}          {\ifmmode\mathrm{\tauup}\else$\mathrm{\tauup}$\fi}

\newcommand*{\egamma}         {E_{\mathrm{\gammaup}}}
\newcommand*{\photon}         {\ifmmode{\gammaup}\else${\gammaup}$\fi}
\newcommand*{\positron}       {\ifmmode{\mathrm{e}^+}\else${\mathrm{e}^+}$\fi}
\newcommand*{\electron}       {\ifmmode{\mathrm{e}}\else${\mathrm{e}}$\fi}
\newcommand*{\epositron}      {{E_\mathrm{e^+}}}
\newcommand*{\ppositron}      {{p_\mathrm{e^+}}}
\newcommand*{\tpositron}      {{t_\mathrm{e^+}}}

\newcommand*{\tgamma}         {{t_{\mathrm{\gammaup}}}}
\newcommand*{\tegamma}        {{t_{\mathrm{e^+ \gammaup}}}}
\newcommand{\tg}{{\ifmmode t_{\gammaup_1\mathrm{e}^+}\else$t_{\gammaup_1\mathrm{e}^+}$\fi}}
\newcommand*{\tgg}{{\ifmmode t_{\gammaup\gammaup}\else$t_{\gammaup\gammaup}$\fi}}
\newcommand*{\Thetaegamma}    {{\Theta_{\mathrm{e}^+ \gammaup}}}

\newcommand*{\thetaegamma}    {{\theta_{\mathrm{e}^+ \gammaup}}}
\newcommand*{\phiegamma}      {{\phi_{\mathrm{e}^+ \gammaup}}}
\newcommand*{\thetae}         {{\theta_\mathrm{e^+}}}
\newcommand*{\phie}           {{\phi_\mathrm{e^+}}}
\newcommand*{\thetagamma}     {{\theta_\mathrm{\gammaup}}}
\newcommand*{\phigamma}       {{\phi_\mathrm{\gammaup}}}

\newcommand{\meg}{\ifmmode{\muup \to e \gammaup}\else$\mathrm{\muup \to e \gammaup}$\fi}
\newcommand{\megp}{\ifmmode{\muup^+ \to \mathrm{e}^+ \gammaup}\else$\mathrm{\muup^+ \to e^+ \gammaup}$\fi}
\newcommand{\michel}{\ifmmode{\muup^+ \to e^+ \nuup\bar{\nuup}}\else$\mathrm{\muup^+ \to e^+ \nuup\bar{\nuup}}$\fi}
\newcommand{\radiative}{\ifmmode{\muup^+ \to \mathrm{e}^+\nuup\bar{\nuup}\gammaup} \else$\mathrm{\muup^+ \to e^+ \nuup\bar{\nuup}\gammaup}$\fi}
\newcommand{\conv}{\ifmmode{\muup^- \to e^-}\else$\mathrm{\muup^- \to e^-}$\fi}
\newcommand{\convN}{\ifmmode{\muup^-N \to e^-N}\else$\mathrm{\muup^-N \to e^-N}$\fi}
\newcommand{\mute}{\ifmmode{\muup \to 3e}\else $\mathrm{\muup \to 3e}$\fi}
\newcommand{\mutec}{\ifmmode{\muup^+ \to e^+e^+e^-}\else $\mathrm{\muup^+ \to e^+e^+e^-}$\fi}
\newcommand{\aif}{\ifmmode\mathrm{e}^+ \mathrm{e}^- \to \gammaup\gammaup \else$\mathrm{e}^+ \mathrm{e}^- \to \gammaup \gammaup$\fi}
\newcommand{\teg}{\ifmmode{\tauup \to e \gammaup} \else$\mathrm{\tauup \to e \gammaup}$\fi}
\newcommand{\tmg}{\ifmmode{\tauup \to \gammaup} \else$\mathrm{\tauup \to \muup \gammaup}$\fi}
\newcommand{\tmueg}{\ifmmode{\mathrm\tauup \to \ell \gammaup}\else$\mathrm{\tauup \to \ell \gammaup}$\fi}
\newcommand{\tautl}{\ifmmode{\mathrm\tauup \to 3\ell} \else$\mathrm\tauup \to 3\ell$\fi}

\newcommand*{\BR}     { {\cal B} }
\newcommand*{\xpos}          {x_\mathrm{e^+}}
\newcommand*{\ypos}          {y_\mathrm{e^+}}
\newcommand*{\zpos}          {z_\mathrm{e^+}}

\newcommand*{\ugamma}         {u_{\gammaup}}
\newcommand*{\vgamma}         {v_{\gammaup}}
\newcommand*{\wgamma}         {w_{\gammaup}}

\newcommand*{\nsig}           {N_{\rm sig}}
\newcommand*{\nrd}            {N_{\rm RMD}}

\newcommand*{\nacc}            {N_{\rm ACC}}
\newcommand*{\xt}            {x_{\rm T}}

\newcommand*{\sens}     { {\cal S}_{90}}
\newcommand*{\ul}     { {\cal B}_{90}}
\newcommand*{\bestfit}     { {\cal B}_\mathrm{fit}}
\newcommand*{\rsig}           {R_{\rm sig}} 

\newcommand*{\mathtentative}{}
\def\mathtentative#1#{\mathcoloraux{#1}}
\newcommand*{\mathcoloraux}[3]{%
  \protect\leavevmode
  \begingroup
    \color#1{#2}#3%
  \endgroup
}

\journalname{Eur. Phys. J. C} 

\begin{document}


\title{A search for \megp\ with the first dataset of the MEG~II experiment}

\author{The MEG~II collaboration}
\newcommand*{\INFNPi}{INFN Sezione di Pisa$^{a}$; Dipartimento di Fisica$^{b}$ dell'Universit\`a, Largo B.~Pontecorvo~3, 56127 Pisa, Italy}
\newcommand*{\INFNGe}{INFN Sezione di Genova$^{a}$; Dipartimento di Fisica$^{b}$ dell'Universit\`a, Via Dodecaneso 33, 16146 Genova, Italy}
\newcommand*{\INFNPv}{INFN Sezione di Pavia$^{a}$; Dipartimento di Fisica$^{b}$ dell'Universit\`a, Via Bassi 6, 27100 Pavia, Italy}
\newcommand*{\INFNRm}{INFN Sezione di Roma$^{a}$; Dipartimento di Fisica$^{b}$ dell'Universit\`a ``Sapienza'', Piazzale A.~Moro, 00185 Roma, Italy}
\newcommand*{\INFNNa}{INFN Sezione di Napoli, Via Cintia, 80126 Napoli, Italy}
\newcommand*{\INFNLe}{INFN Sezione di Lecce$^{a}$; Dipartimento di Matematica e Fisica$^{b}$ dell'Universit\`a del Salento, Via per Arnesano, 73100 Lecce, Italy}
\newcommand*{\ICEPP} {ICEPP, The University of Tokyo, 7-3-1 Hongo, Bunkyo-ku, Tokyo 113-0033, Japan }
\newcommand*{\Kobe} {Kobe University, 1-1 Rokkodai-cho, Nada-ku, Kobe, Hyogo 657-8501, Japan}
\newcommand*{\UCI}   {University of California, Irvine, CA 92697, USA}
\newcommand*{\KEK}   {KEK, High Energy Accelerator Research Organization, 1-1 Oho, Tsukuba, Ibaraki 305-0801, Japan}
\newcommand*{\PSI}   {Paul Scherrer Institut PSI, 5232 Villigen, Switzerland}
\newcommand*{\Waseda}{Research Institute for Science and Engineering, Waseda~University, 3-4-1 Okubo, Shinjuku-ku, Tokyo 169-8555, Japan}
\newcommand*{\BINP}  {Budker Institute of Nuclear Physics of Siberian Branch of Russian Academy of Sciences, 630090 Novosibirsk, Russia}
\newcommand*{\JINR}  {Joint Institute for Nuclear Research, 141980 Dubna, Russia}
\newcommand*{\ETHZ}  {Institute for Particle Physics and Astrophysics, ETH Z\" urich, 
Otto-Stern-Weg 5, 8093 Z\" urich, Switzerland}
\newcommand*{\NOVS}  {Novosibirsk State University, 630090 Novosibirsk, Russia}
\newcommand*{\NOVST} {Novosibirsk State Technical University, 630092 Novosibirsk, Russia}
\newcommand*{\ScuolaPi}{Scuola Normale Superiore, Piazza dei Cavalieri 7, 56126 Pisa, Italy}
\newcommand*{\INFNLNF}{\textit{Present Address}: INFN, Laboratori Nazionali di Frascati, Via 
E. Fermi, 40-00044 Frascati, Rome, Italy}
\newcommand*{\Liverpool}{Oliver Lodge Laboratory, University of Liverpool, Liverpool, L69 7ZE, United Kingdom}

\date{Received: date / Accepted: date}

\author{
The MEG~II collaboration\\\\
        K.~Afanaciev~\thanksref{addr12} \and
        A.~M.~Baldini\thanksref{addr1}$^{a}$ \and
        S.~Ban~\thanksref{addr10} \and
        V.~Baranov~\thanksref{addr12}\thanksref{e5}  \and
        H.~Benmansour\thanksref{addr1}$^{ab}$ \and
        M.~Biasotti~\thanksref{addr3}$^{a}$ \and
        G.~Boca~\thanksref{addr4}$^{ab}$ \and        P.~W.~Cattaneo~\thanksref{addr4}$^{a}$\thanksref{e1} \and
        G.~Cavoto~\thanksref{addr5}$^{ab}$ \and
        F.~Cei~\thanksref{addr1}$^{ab}$ \and
        M.~Chiappini~\thanksref{addr1}$^{ab}$ \and
        G.~Chiarello~\thanksref{addr6}$^{a}$\thanksref{e2} \and
        A.~Corvaglia~\thanksref{addr6}$^{a}$ \and
        F.~Cuna~\thanksref{addr6}$^{ab}$\thanksref{e3} \and
        G.~Dal~Maso\thanksref{addr2,addr16} \and
        A.~De~Bari~\thanksref{addr4}$^{a}$ \and
        M.~De~Gerone~\thanksref{addr3}$^{a}$ \and
        L.~Ferrari~Barusso~\thanksref{addr3}$^{ab}$ \and
        M.~Francesconi~\thanksref{addr17} \and 
        L.~Galli~\thanksref{addr1}$^{a}$ \and
        G.~Gallucci~\thanksref{addr3}$^{a}$ \and
        F.~Gatti~\thanksref{addr3}$^{ab}$ \and
        L.~Gerritzen~\thanksref{addr10}  \and
        F.~Grancagnolo~\thanksref{addr6}$^{a}$ \and
        E.~G.~Grandoni~\thanksref{addr1}$^{ab}$ \and 
        M.~Grassi~\thanksref{addr1}$^{a}$ \and 
        D.~N.~Grigoriev~\thanksref{addr7,addr8,addr9} \and
        M.~Hildebrandt~\thanksref{addr2} \and
        K.~Ieki~\thanksref{addr10}  \and
        F.~Ignatov~\thanksref{addr15} \and
        F.~Ikeda~\thanksref{addr10}  \and
        T.~Iwamoto~\thanksref{addr10}  \and
        S.~Karpov~\thanksref{addr7,addr9} \and
        P.-R.~Kettle~\thanksref{addr2} \and
        N.~Khomutov~\thanksref{addr12} \and
        S.~Kobayashi~\thanksref{addr10}  \and
        A.~Kolesnikov~\thanksref{addr12}  \and
        N.~Kravchuk~\thanksref{addr12}  \and
        V.~Krylov~\thanksref{addr12} \and
        N.~Kuchinskiy~\thanksref{addr12}  \and
        W.~Kyle~\thanksref{addr11} \and
        T.~Libeiro~\thanksref{addr11} \and  
        V.~Malyshev~\thanksref{addr12}  \and
        A.~Matsushita~\thanksref{addr10}  \and
        M.~Meucci~\thanksref{addr5}$^{ab}$ \and   
        S.~Mihara~\thanksref{addr13}  \and
        W.~Molzon~\thanksref{addr11} \and
        Toshinori~Mori~\thanksref{addr10}  \and
        M.~Nakao~\thanksref{addr10} \and 
        D.~Nicol\`o~\thanksref{addr1}$^{ab}$ \and
        H.~Nishiguchi~\thanksref{addr13}  \and
        A.~Ochi~\thanksref{addr14}  \and
        S.~Ogawa~\thanksref{addr10}  \and
        R.~Onda~\thanksref{addr10}  \and
        W.~Ootani~\thanksref{addr10}  \and
        A.~Oya~\thanksref{addr10} \and
        D.~Palo~\thanksref{addr11} \and
        M.~Panareo~\thanksref{addr6}$^{ab}$ \and
        A.~Papa~\thanksref{addr1}$^{ab}$\thanksref{addr2} \and
        V.~Pettinacci~\thanksref{addr5}$^{a}$ \and
        A.~Popov~\thanksref{addr7,addr9} \and
        F.~Renga~\thanksref{addr5}$^{a}$ \and
        S.~Ritt~\thanksref{addr2} \and
        M.~Rossella~\thanksref{addr4}$^{a}$ \and
        A.~Rozhdestvensky~\thanksref{addr12}  \and
        P.~Schwendimann~\thanksref{addr2} \and
        K.~Shimada~\thanksref{addr10} \and
        G.~Signorelli~\thanksref{addr1}$^{a}$ \and
        M.~Takahashi~\thanksref{addr14}  \and
        G.F.~Tassielli~\thanksref{addr6}$^{ab}$\thanksref{e4} \and
        K.~Toyoda~\thanksref{addr10} \and
        Y.~Uchiyama~\thanksref{addr10,addr14} \and
        M.~Usami~\thanksref{addr10} \and
        A.~Venturini~\thanksref{addr1}$^{ab}$ \and
        B.~Vitali~\thanksref{addr1}$^{a,}$\thanksref{addr5}$^{b}$ \and
        C.~Voena~\thanksref{addr5}$^{ab}$ \and   
        K.~Yamamoto~\thanksref{addr10}  \and
        K.~Yanai~\thanksref{addr10} \and
        T.~Yonemoto~\thanksref{addr10}  \and
        K.~Yoshida~\thanksref{addr10} \and
        Yu.V.~Yudin~\thanksref{addr7,addr9} 
}

\institute{\JINR   \label{addr12}
           \and
             \INFNPi \label{addr1}
           \and
             \ICEPP \label{addr10}
           \and
             \INFNGe \label{addr3}
            \and
             \INFNPv \label{addr4}
           \and
             \INFNRm \label{addr5}
           \and
             \INFNLe \label{addr6} 
           \and
             \PSI \label{addr2}
            \and
             \ETHZ \label{addr16}
           \and
             \INFNNa \label{addr17} 
            \and
             \BINP   \label{addr7}
           \and
             \NOVST  \label{addr8}
           \and
             \NOVS   \label{addr9}
         \and
             \Liverpool  \label{addr15}
           \and
             \UCI    \label{addr11}
           \and
             \KEK    \label{addr13}
           \and
             \Kobe    \label{addr14}
}

\thankstext[*]{e1}{Corresponding author: paolo.cattaneo@pv.infn.it} 
\thankstext[**]{e2}{Presently at Department of Engineering, University of Palermo, Viale delle Scienze,
Building 9, 90128 Palermo, Italy} 
\thankstext[***]{e3}{Presently at INFN Sezione di Bari, Via Giovanni Amendola, 173, 70126, Bari, Italy} 
\thankstext[****]{e4}{Presently at Dipartimento di Medicina e Chirurgia, Università LUM “Giuseppe
Degennaro”, 70010, Casamassima, Bari, Italy}
\thankstext[$\dagger $]{e5}{Deceased} 

\maketitle 

\begin{abstract}
The MEG~II experiment, based at the Paul Scherrer Institut in Switzerland, reports 
the result of a search for the decay \megp\ from data taken in the first physics run in 2021. 
No excess of events over the expected background is
observed, yielding an upper limit on the branching ratio of
${\cal B} (\megp) < \num{7.5e-13}$ (\SI{90}{\percent} C.L.).
The combination of this result and the limit obtained by MEG gives 
${\cal B} (\megp) < \num{3.1e-13}$ (\SI{90}{\percent} C.L.), which
is the most stringent limit to date. A 
ten-fold larger sample of data is being collected during the years 2022--2023, and data-taking will continue in the coming years.
\end{abstract}


\tableofcontents 
\section{Introduction}

In the standard model (SM) of particle physics, charged lepton flavour-violating 
(CLFV) processes  are basically forbidden with only extremely small 
branching ratios ({\color{black} \num{\sim e-54} \cite{calibbi_2018}}) when
accounting for non-zero neutrino mass differences and mixing angles. Therefore, such decays are free from SM physics backgrounds and a 
positive signal would be unambiguous evidence for physics beyond the SM. 
Several SM extensions predict CLFV decays at measurable rates, and the 
channel \megp\ is particularly sensitive to new physics.
Reviews of the theoretical expectations and experimental 
status are provided in \cite{calibbi_2018,Mihara:2013zna}.

The MEG collaboration searched for the \megp\ decay at the Paul Scherrer 
Institut (PSI) in Switzerland in the period 2008–2013, improving the previous limit on the branching ratio by more than an order of magnitude, 
down to ${\cal B} (\megp) < 4.2 \times  10^{-13}$ (90\% C.L.) 
\cite{baldini_2016}.
A detailed report of the MEG experiment's motivation and design criteria 
is available in \cite{megdet} and references therein.

In this paper, we report the first result of the MEG~II experiment, an upgrade of MEG aiming to improve the sensitivity to the \megp\ branching ratio by one order of magnitude within the next few years. 

\section{Signal and background}


The event signature is given by a \photon-ray and a positron, forming a pair with the kinematic features of a two-body decay at rest. In particular, the positron and \photon-ray are emitted at the same time $\tpositron = \tgamma$ 
($\tegamma \equiv \tgamma - \tpositron = 0$), and with the same energy, 
$\epositron\ \approx \egamma\ \approx \mathrm{m_\mu} \mathrm{c}^2/2 \approx \SI{52.83}{\MeV}$ (the 
positron mass is negligible, given the detector resolutions), in opposite directions:
\begin{eqnarray}
\nonumber \thetaegamma &\equiv& (\pi - \thetae) - \thetagamma = 0 \; ,\\
\nonumber \phiegamma &\equiv& (\pi + \phie) - \phigamma = 0  \pmod {2\pi} \; ,
\end{eqnarray}
where $\phie$ and $\thetae$ ($\phigamma$ and $\thetagamma$) are the azimuthal and polar angles of the positron (\photon-ray).


The background has two components: one from the 
radiative muon decay (RMD) \radiative\ and one from the 
accidental superposition of an energetic positrons from the 
standard muon Michel decay with a high energy \photon-ray from RMD, 
positron--electron annihilation-in-flight or bremsstrahlung (ACC).
For $\egamma > \SI{51.5}{\MeV}$, the \photon-rays from annihilation-in-flight
dominate.
At the MEG~II data taking rate in 2021, more than \SI{90}{\percent} of collected events  with 
$\egamma > \SI{48}{\MeV}$ are from the ACC background.


The ACC background is characterised  by wide distributions in 
$\epositron$ and $\egamma$, dropping to zero at the kinematic endpoint at \SI{52.83}
{\MeV}, and wide distributions in the relative angles, almost flat around 
$\phiegamma = \thetaegamma = 0$. The distribution of $\tegamma$ is flat because the
positron and the \photon-ray originate from the decays of different muons.

The RMD background is characterised by an anticorrelated distribution 
of $\epositron$ and $\egamma$, also dropping to zero at the kinematic endpoint. The 
angular distribution is peaked with positron and \photon-ray aligned, while the 
back-to-back configuration is highly suppressed. The distribution of $\tegamma$ is peaked at zero.

\section{The MEG~II experiment}

\begin{figure}[!htb]
\centering
\vspace{-0.5cm}
  \includegraphics[width=0.45\textwidth,angle=0] {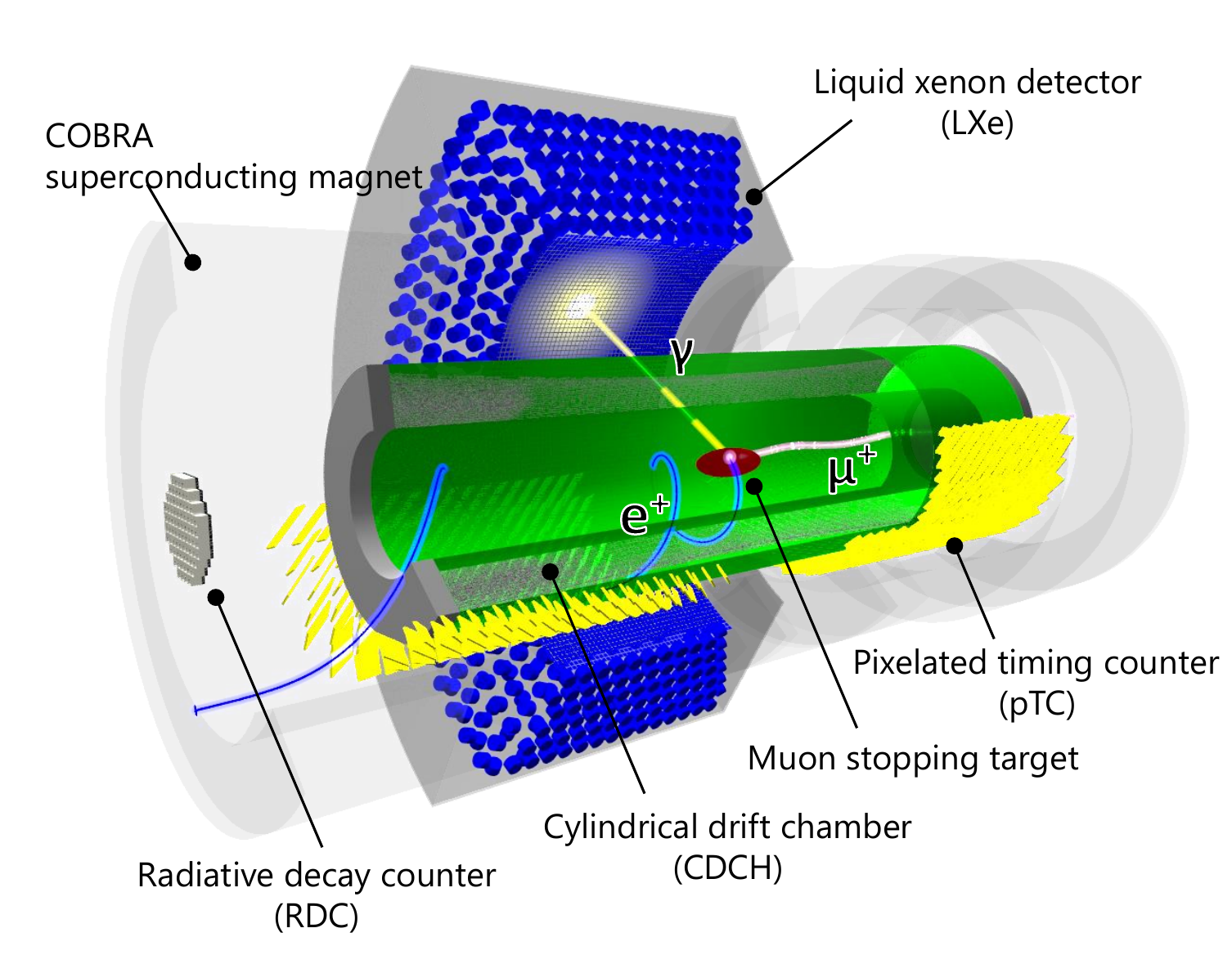}
  \caption{A sketch of the MEG~II detector with a simulated \megp\ event.}
 \label{meg2det}
\end{figure}


The MEG~II detector, located at the $\piup$E5 beam line at PSI, is designed to measure with high 
precision the positron and \photon-ray kinematics and the relative 
production time of the two particles,
coping with high \muonp\ stopping rates up to {\color{black} $R_\mu = \SI{5e7}{\per\second}$}.
A detailed description of the MEG~II 
detector and its performance is in \cite{megII-det}, and a sketch is shown in~\fref{meg2det}. {\color{black} A right-handed, Cartesian coordinate system is adopted, with the $z$ axis along the beam direction and the $y$-axis vertical and pointing upward.}

%
Briefly, a spectrometer is built inside a Constant Bending RAdius (COBRA) 
superconducting magnet, generating a gradient magnetic field with maximum intensity \SI{1.27}{\tesla} so as
to contain the positrons emitted by \megp\ decays in a thin muon stopping target 
at the centre within the bore of the magnet and sweep them quickly outside.

{\color{black}
The target is an elliptical foil (\SI{270}{\milli\meter} long and 
\SI{66}{\milli\meter} high)
with \SI{174\pm 20}{\micro\meter} average thickness.
The direction normal to The target foil normal lies on 
the $(x,z)$ plane and forms an angle of
\SI{75.0 \pm 0.1}{\degree} with respect to the beam axis ($z$-axis). This design maximise the muon stopping probability minimising the material crossed by the outgoing particles.
}

The spectrometer is instrumented with a single-volume, gaseous cylindrical drift chamber (CDCH)~\cite{Chiappini:2023egy}, and two sectors of scintillating tiles forming the pixelated timing counter (pTC) \cite{Nishimura2020}, all placed inside the bore of COBRA.

The CDCH is a \SI{1.93}{\meter}-long, low-mass cylindrical volume, filled 
with a helium--isobutane gas mixture with the addition of small percentages of oxygen and isopropyl alcohol to avoid current spikes. It has nine concentric 
layers of gold-plated tungsten sense wires, arranged in a stereo configuration with two views. The drift cells, delimited by silver-plated aluminium wires, have a nearly square  shape, with sides ranging from 
\SI{5.8}{\milli\meter} in the central part of the innermost layer to \SI{8.7}{\milli\meter} at the end plates of the outermost layer.

The pTC consists of two semicylindrically 
shaped sectors, one located upstream of the target
and the other downstream, designed to provide precise measurements of the positron timing.
Each sector consists of 256 scintillator tiles, each read out by two arrays of six SiPMs. A signal positron hits on average
\num{\sim 9} tiles, which provide independent measurements of the positron crossing timing with a resolution of \SI{\sim 100}{\pico\second}. The overall time resolution is 
$\sigma_{\tpositron,\rm{pTC}} = \SI{43}{\pico\second}$.

A liquid xenon detector (LXe), located outside of COBRA, consists of a homogeneous volume (\SI{900}{\litre}) of liquid xenon viewed by 4092 Multi-Pixel Photon Counters (MPPCs), located on the front face~\cite{IEKI2019148},
and 668 UV-sensitive photomultiplier tubes (PMTs), all submerged in the liquid. This detector subtends the region $\phi_\photon \in \left( \frac{2}{3} \pi, \frac{4}{3} \pi\right)$ and
$|\cos \theta_\photon | < 0.35$, corresponding to
\SI{\sim 11}{\percent} of the solid angle, determining the geometrical acceptance of MEG~II for \megp\ decays. {\color{black} The efficiencies given below refer to this acceptance.}

The radiative decay counter (RDC) is a novel detector, located downstream
and centred on axis, designed to identify the ACC events with an
RMD-originated high-energy \photon-ray by tagging a low energy positron in coincidence. The RDC consists of a scintillating plastic detector to measure the positron timing and a LYSO crystal calorimeter to measure the energy.

The highly integrated trigger and data acquisition system called WaveDAQ~\cite{francesconi2023wavedaq} is based on WaveDREAM modules. 
They make use of the DRS4 chip to digitise the signals from the detectors at 1.4 GSPS (1.2 GSPS for CDCH) sampling speed. 
The waveforms are then analysed offline to extract time and amplitude information with high precision.

The trigger for \megp\ events is based on the online estimate of $\egamma$ with the LXe 
detector, on the relative time between the positron and the $\photon$-ray $T_{\positron\photon}$ measured by the LXe and the pTC and on the direction match measured by the same detectors. 
{\color{black}
Trigger parameters had to be tuned during data taking. In addition, they depend on $R_\mu$ and must be recalculated separately for each beam rate.
The time needed to reach a satisfactory tuning limited the average trigger efficiency to 
$\varepsilon_\mathrm{TRG} = \SI{80(1)}{\percent}$ in the analysis of the dataset presented here. 
On the basis of the past experience we expect for the 
following runs $\varepsilon_\mathrm{TRG}$ to be close to \SI{95}{\percent}.
}

The apparatus requires constant monitoring and calibrations.
{\color{black} Dedicated} instrumentation has been developed, such as: 
{\color{black} dedicated runs with a $\pi^-$ beam producing photons through the charge-exchange (CEX) reaction $\pi^- + p \to \pi^0(\gamma \gamma) + n$}, a Cockroft--Walton accelerator (CW), a neutron generator, LED and $\alpha$-particles 
submerged in liquid xenon for LXe detector 
energy calibration; a laser system for pTC timing calibration; photo cameras for measuring precisely the target position  \cite{calibration_cw,Signorelli:2015,laserpTC,Palo:2019gzw,Cavoto:2020etw}.

For the LXe, a local system of curvilinear coordinates $(u,v,w)$ is also used, where $u$ and $v$ are tangent to the cylindrical inner surface of the calorimeter (with $u$ parallel to $z$) and $w$ is the depth inside the liquid xenon fiducial volume.

\section{Event reconstruction}

In each event, positron and \photon-ray candidates are described by five observables: $\egamma$, $\epositron$, $\phiegamma$, $\thetaegamma$ and $\tegamma$.



The positron kinematics is reconstructed by tracking the trajectory of the particle in the magnetic spectrometer and extrapolating it back to the muon stopping target~\cite{CDCHPerfor}. 

Electronic waveforms are collected and digitised on both sides of the sense wires in the 
CDCH, digitally filtered to suppress the noise and analysed with both conventional 
and machine-learning techniques to extract the time and the induced charge of the ionisation signals (\emph{hits}) ~\cite{CDCHPerfor}.

Hits are combined into tracks by a track-following 
algorithm, which starts from clusters of near wires in the external layers of the chamber and propagates them through the detector, adding new hits with the help of a Kalman filter. 

In parallel, scintillation signals in the tiles of the pTC are 
reconstructed from the SiPM waveforms and combined in clusters of close tiles, from 
which an estimate of the positron time is extracted. The combination of the CDCH hit 
times with the positron time in the pTC allows for  precise determination of the drift 
distance of the ionisation electrons in the CDCH cell and hence the distance of 
closest approach (DOCA) of the positron trajectory to the wires. In this procedure, 
an innovative machine-learning procedure is used to extract the DOCA, using the full 
signal waveforms as inputs~\cite{Docauci}, instead of the drift times extracted with 
conventional approaches. The typical precision of the DOCA reconstruction is about 
\SI{115}{\micro\meter}. 

Once a track candidate is built and the DOCA of each hit has been precisely determined, a Kalman filter complemented by a deterministic annealing filter~\cite{DAF}, including the effect of the positron interactions with the detector material, is used to fit the track. The track is finally extrapolated to the intermediate plane of the muon stopping target, where the positron position $(\xpos, \ypos, \zpos)$ and momentum$(\ppositron,\thetae,\phie)$ are determined. It is also propagated to the corresponding pTC cluster, and the total trajectory length $l_{\positron}$ from the target to pTC cluster is measured, with a resolution $O(\SI{10}{\pico\second})$. The positron time $\tpositron$ is determined as the pTC cluster time minus the positron time of flight $l_{\positron}/\mathrm{c}$. 

The efficiency of the track reconstruction in the CDCH is {\color{black} $\varepsilon_{\positron,\rm{CDCH}}= (74.0\pm 1.5 \pm 4.0_{R_\mu})\% $}
at $R_\mu = \SI{3e7}{\per\second}$, 
mainly limited by the pileup of multiple tracks in the same event and hence deteriorating with 
increasing beam rates. The second uncertainty is due to a systematic error on the measurement of $R_\mu$ and is fully correlated among maesurements of $\varepsilon_{\positron,\rm{CDCH}}$ performed at different $R_\mu$.
Including the {\color{black} pTC acceptance and efficiency for signal positrons},
{\color{black} $\varepsilon_{\positron,\rm{pTC}}= \SI{91(2)}{\percent}$}, the positron reconstruction efficiency results {\color{black} $\varepsilon_{\positron} = (67.0\pm 1.5 \pm 4.0_{R_\mu})\% $
}.

The tracking efficiency {\color{black} $\varepsilon_{\positron,\rm{CDCH}}$ decreases from 
$(77.0\pm 1.5 \pm 4.0_{R_\mu})\% $
to 
$(66.0\pm 1.5 \pm 4.0_{R_\mu})\% $
when $R_\mu$ increases from \SI{2e7}{\per\second} to \SI{5e7}{\per\second}. This trend dominates the dependence of $\varepsilon_{\positron}$ on $R_\mu$.}

The \photon-rays are measured in the LXe detector from the combination of the 
individual MPPC and PMT signals. The digitised waveforms are filtered by subtracting 
average noise templates extracted from pedestal runs, where events are collected 
without beam on target and with a periodic trigger. Then, the charge collected in 
each sensor is measured by integrating the waveform in a \SI{150}{\nano\second} window around the 
expected signal time and converted into the number of scintillation photons by means 
of gains and quantum efficiencies (for PMTs) or photon detection efficiencies (for 
MPPCs) extracted from dedicated calibrations. 

For the measurement of the first conversion 
position of the incident \photon-ray ($\ugamma$, $\vgamma$, $\wgamma$), a $\chi^2$ is minimised, 
which compares the number of observed photons in 
the MPPCs to the number of expected photons for 
\photon-ray's converting in a given position. Similarly, once the position 
of the \photon-ray conversion is known, the conversion time $t_\mathrm{LXe}$ is determined by minimising a 
$\chi^2$ based on the expected and observed arrival times of the scintillation 
photons to both PMTs and MPPCs. Finally, the energy of the \photon-ray is determined by 
summing the number of photons in all sensors and converting it into an energy value 
by means of several correction factors. They account for the average light yield of 
the LXe, the position-dependent photosensor coverage and light detection efficiency, 
the evolution of the sensor response during the run, and residual non-uniformities in 
the response of the detector. 
The overall efficiency for signal \photon-rays is 
$\varepsilon_\photon = \SI{62(2)}{\percent}$.

The direction of the \photon-ray cannot be precisely 
measured in the LXe detector. Consequently, a direct reconstruction of the positron-\photon-ray
relative angles is not possible, and an indirect approach is used: the positron 
position $(x_\positron, y_\positron, z_\positron)$ at the target is assumed to be the muon decay point and 
hence also the production point of the \photon-ray for signal events. 
Therefore, the \photon-ray direction $(\thetagamma,\phigamma)$ is taken as the one 
joining the positron 
position at the target and the detection point in the LXe detector. 

The time of flight from 
the supposed muon decay point to the \photon-ray detection point is subtracted from the 
conversion time to determine the \photon-ray production time $\tgamma$. 
The resolution on $\tegamma$ is dominated by the time resolution of the LXe detector ($\sigma_{t_{\photon,\rm{LXe}}} = \SI{65}{\pico\second}$).

The RDC measures the time $t_{\positron,\rm{RDC}}$ and energy loss $E_{\positron,\rm{RDC}}$ 
of a low-energy positron in coincidence with a high-energy $\photon$-ray measured in the LXe detector.
The distributions of $t_{\positron,\rm{RDC}} - t_{\photon,\rm{LXe}}$ 
and $E_{\positron,\rm{RDC}}$ differ between signal (the former is flat and the latter is peaked at high energy) and 
ACC background with an RMD-originated \photon-ray (the former is peaked around zero while the latter is peaked at low energy), providing additional discriminating power.

Details on the reconstruction algorithms and calibration procedures can be found 
in~\cite{megII-det}. \tref{perf} shows the performance achieved on the 2021 dataset, in terms of resolutions and efficiencies.

\begin{table}
\caption{ \label{perf}Resolutions (Gaussian $\sigma$) and efficiencies of the MEG~II experiment, measured at $R_\mu = \SI{3e7}{\per\second}$.}
\centering
\newcommand{\minu}{\hphantom{$-$}}
\newcommand{\cc}[1]{\multicolumn{1}{c}{#1}}
\begin{tabular}{@{}lll}
\hline
  {\bf Resolutions }  & \minu \\ 
\hline\noalign{\smallskip}
$\epositron$ (\unit{keV})  & \minu 89 \\
$\phie,\thetae$ (\unit{mrad})    & \minu 4.1/7.2 \\
$\ypos,\zpos$ (\unit{mm})   & \minu 0.74/2.0  \\
$\egamma$(\%)  ($\wgamma\SI{<2}{\cm}$)/($\wgamma\SI{>2}{\cm}$)  & \minu {2.0/1.8} \\
$\ugamma, \vgamma, \wgamma,$ (\unit{mm})
& \minu {2.5/2.5/5.0} \\
$\tegamma$ (\unit{ps}) & \minu 78 \\
\hline
{\bf  Efficiencies (\%)} & \\ 
\hline
$\varepsilon_{\photon}$     & \minu 62 \\
$\varepsilon_{\positron}$   & \minu 67 \\
$\varepsilon_\mathrm{TRG}$  & \minu 80  \\
\hline
\end{tabular}
\end{table}

\section{Analysis}
\subsection{Overview}
The data analysed in this work were collected in the year 2021 during the first, seven-week-long physics run of MEG~II, with a total DAQ livetime of \SI{2.9e6}{\second}. The data-taking was performed at four
different beam intensities ($R_\mu = \num{2e7}$, 
\num{3e7}, \num{4e7}, $\SI{5e7}{\per\second}$) in five different periods of time (in two of them, the beam {\color{black} intensity} was $R_\mu = \SI{3e7}{\per\second}$) to study 
the beam rate dependence of the detector performance.
A total of \num{1.04e14} \muonp\ were stopped on the target. The fractions of the integrated \muonp\ on target for the above intensities are (0.13, 0.41, 0.20, 0.26), respectively. The \megp\ trigger rates went from $\SI{\sim 4}{\Hz}$ to $\SI{\sim 20}{\Hz}$. The size of the \megp\ trigger sample was \num{\sim 2.e7}.

As in the MEG experiment~\cite{baldini_2016}, 
an unbinned maximum likelihood technique is applied in the analysis region defined by $\SI{48}{\MeV} < \egamma < \SI{58}{\MeV}$, $\SI{52.2}{\MeV} < \epositron < \SI{53.5}{\MeV}$, $|\tegamma|< \SI{0.5}{\nano\second}$, $|\phiegamma|< \SI{40}{\milli\radian}$ and $|\thetaegamma| < \SI{40}{\milli\radian}$. 

This approach is adopted for a blind analysis:
the events 
in a ``blinding box'' defined as $48.0<\egamma<\SI{58.0}{\MeV}$ and $|\tegamma|<\SI{1}{\ns}$, which 
includes the analysis region, are initially hidden; only once the probability density functions (PDFs) of observables used to discriminate signal from background are ready to build a likelihood function $\mathcal{L}(\nsig)$, the hidden data are released and used to extract a confidence interval for the expected number of signal events, $\nsig$.


All necessary studies on the background, including the construction of the PDFs, are done in side-bands outside the analysis region. The regions defined by $\SI{1}{\nano\second} < |\tegamma| < \SI{3}{\nano\second}$ are called ‘‘time side-bands’’, and are used to study the ACC background. The region defined by $\SI{45}{\MeV} < \egamma < \SI{48}{\MeV}$  is called ‘‘$\egamma$ side-band’’. It includes RMD events peaking at $\tegamma = 0$, and is used to extract the 
$\tegamma$ PDF for both RMD and signal events.

\subsection{Confidence interval}

The construction of the confidence interval for the number of signal $\nsig$ events is based on the Feldman--Cousins prescription~\cite{feldman_1998}, 
with the profile likelihood ratio ordering \cite{PDBook_2014}.
The profile likelihood ratio $\lambda_p$ is defined as
\begin{eqnarray}\label{eq:test_statistic}
\lambda_{p}(\nsig) &=&
\left\{
\begin{array}{lll}
\frac{{\mathcal L}(\nsig      ,  \hat{\hat{\vector{\theta}}}(\nsig))}
     {{\mathcal L}(0          ,  \hat{\hat{\vector{\theta}}}(0    ))}
 & {\rm if } & \hat{N}_{\rm sig} <   0\\
\frac{{\mathcal L}(\nsig      ,  \hat{\hat{\vector{\theta}}}(\nsig))}
     {{\mathcal L}(\hat{N}_{\rm sig},       \hat{\vector{\theta}}        )}
 & {\rm if } & \hat{N}_{\rm sig} \ge 0 \; ,
\end{array} \right.\nonumber 
\end{eqnarray}
where ${\vector{\theta}}$ is a vector of nuisance parameters; $\hat{N}_{\rm sig}$ 
and $\hat{\vector{\theta}}$
are the values of
$\nsig$ and $\vector{\theta}$ that maximise the likelihood; $\hat{\hat{\vector{\theta}}}(\nsig)$
is the value of $\vector{\theta}$ which maximises the likelihood for the specified $\nsig$.

The systematic uncertainties on the PDFs and the normalisation factor 
described in the next section are incorporated with two methods: either 
profiling them as nuisance parameters in the likelihood function or randomly fluctuating the PDFs according to the uncertainties.
The profiling method is generally known to be more robust than the random 
fluctuation method, but it requires CPU-intensive calculations.
It is, therefore, employed only for the uncertainty with the largest contribution, 
which is the detector misalignment, while the others are included by the random fluctuation method.

\subsection{Likelihood function}
The likelihood function is obtained by combining the PDFs for the observables discriminating between signal and background. Besides $\epositron$, $\egamma$, $\tegamma$, $\thetaegamma$ and $\phiegamma$, for events with RDC signals we also exploit the RDC observables ($t_{\positron,\mathrm{RDC}}-t_{\photon,\mathrm{LXe}}$, $E_{\positron,\mathrm{RDC}}$). Moreover, the $\tegamma$ resolution has a relevant dependence on the number of hits in the pTC cluster, $n_\mathrm{pTC}$. In order to take this into account, and considering that $n_\mathrm{pTC}$ has significantly different distributions in signal and background, this quantity is also included in the list of observables.

The extended likelihood function is hence defined as
\begin{subequations}
   \begin{align*}
      &\mathcal{L}(\nsig, \nrd, \nacc, \xt) \nonumber =\\ 
      & \frac{e^{-(\nsig +\nrd+\nacc)}}{N_\mathrm{obs}!} 
      C(\nrd, \nacc, \xt) \times  \nonumber \\
      &\prod_{i=1}^{N_\mathrm{obs}} \bigl(\nsig S(\vec{x_i})+\nrd R(\vec{x_i})+\nacc A(\vec{x_i})\bigr) \; ,
      \label{eq:ExtendedLikelihood}
   \end{align*}
\end{subequations}
where $\vec{x_i} = (\epositron, \egamma, \tegamma, \thetaegamma, \phiegamma, t_{\mathrm{RDC}}-t_{\mathrm{LXe}}, E_{\mathrm{RDC}}, n_\mathrm{pTC})$
is the set of the observables for the $i$-th event;
$S$, $R$ and $A$ are the PDFs for the signal, RMD and ACC background, respectively; $\nsig$, $\nrd$ and $\nacc$ 
are the expected numbers of 
signal, RMD and ACC background events in the analysis region; $\xt$ is a parameter representing the misalignment of 
the muon stopping target;
 $N_\mathrm{obs}$ is the total number of events observed in the analysis region.

In the extraction of the confidence interval for 
$\nsig$, the nuisance parameters are 
${\vector{\theta}} = (\nrd, \nacc, \xt)$, with a 
constraint $C$ applied to their values: 
$\nrd$ and $\nacc$
are Gaussian-constrained by the numbers evaluated in the side-bands and their uncertainties; 
$\xt$ is Gaussian-constrained with its uncertainty. 

Two independent likelihood analyses are performed 
for cross-checking the results with two different types of PDFs: ``per-event PDFs'' 
and ``constant PDFs''.

\subsubsection{Per-event PDF}
The reconstruction performance depends on the detector conditions, 
on the position of the interaction in the detector, 
and other factors changing event by event, such as the 
occurrence of some specific interaction of the particles 
with the detector material. For the ``per-event PDF'' approach, 
the PDF parameters vary on an event-by-event basis to take into 
account these variations. This allows the exploitation of 
the detailed detector information to maximise the 
sensitivity. The PDFs are conditioned by  observables 
that can reflect these variations. 

For the 
\photon-ray PDFs, the resolutions and the background 
spectrum are dependent on the \photon-ray conversion 
position in the LXe detector. For the positron angle, 
vertex position and momentum, an event-by-event 
estimate of the track fit uncertainty can be extracted from 
the covariance matrix of the Kalman filter and used 
to build per-event PDFs. Correlations among the 
positron variables are also taken into account, 
although in this case, instead of extracting the 
parameters from the Kalman covariance matrix, an 
empirical analytic model of the average correlations 
is adopted, taking into account only their $\phie$ 
dependence.

For energies, angles and time, the signal PDFs are 
modelled as Gaussian functions reflecting the measured 
resolutions, with the possible addition of 
tails, according to the results of 
calibrations. The ACC $\epositron$ PDF is the 
combination of the theoretical Michel spectrum with 
acceptance and resolution effects, fitted to data in 
the side-bands~\cite{CDCHPerfor}. The ACC $\egamma$ 
PDF is taken from the Monte Carlo spectrum, with a 
Gaussian smearing and an additional cosmic-ray 
contribution to match the data distribution in the 
side-bands. The ACC angular PDFs are modelled with 
fourth-order polynomials fitted in the side-bands. 
The RMD PDFs are obtained by convolution of the 
theoretical spectra with the experimental 
resolutions. 

The $n_\mathrm{pTC}$ PDFs are taken from the side-bands for the ACC background, and from the Monte Carlo for signal and RMD. 

A special treatment is necessary for the RDC observables, 
because most of the events do not have RDC signals. The RDC PDFs are approximated with binned 2-dimensional distributions, 
with one additional bin reserved for events with no RDC signals. 
The ACC PDFs are extracted from the side-bands, the signal and 
RMD PDFs are extracted from a control sample made of events with signals in the RDC but not in time coincidence with the \photon-ray in the LXe detector.

\subsubsection{Constant PDF}
Another approach for the PDFs' construction uses ``constant PDFs'', and is employed for cross-check purpose.  
The PDFs are constructed with constant parameters by averaging out the temporal variations, the position dependence of the detector response (with the only exception of the conversion depth inside the calorimeter, with different PDFs for $w_\photon < \SI{2}{\centi\meter}$ and $w_\photon > \SI{2}{\centi\meter}$) and the correlations between the observables. The differences in performance at different beam rates are accounted. The relative angle $\Thetaegamma$ between the positron and the \photon-ray, instead of the two separate projections, $\phiegamma$ and $\thetaegamma$ is used.
The RDC observables are not used in this analysis. It makes the analysis simpler and, given the small
statistics of the 2021 dataset, does not deteriorate significantly the sensitivity.

This approach shows worse sensitivity compared to the per-event one, while it's less prone to systematic uncertainties.

\subsection{Normalisation}
\label{sec:Normalisation}
The estimated number of signal events is translated into the branching ratio as 
$\BR(\megp) = \nsig/N_\mu$,
where the normalisation factor $N_\mu$ is the number of effectively measured muon decays in the experiment. $N_\mu$ is evaluated with two independent methods: the number of Michel positrons counted with a dedicated trigger and the number of RMD events measured in the energy side-band~\cite{baldini_2016}.
In both methods, the normalisation dataset is collected 
in parallel with the physics data-taking, such to account possible variations of the detector condition and the instantaneous muon beam intensity.
Both methods give consistent normalisation factors, 
yielding the combined result $N_\mu = (2.64 \pm 0.12) \times 10^{12}$.
%

\subsection{Results}

The following results refer (unless otherwise specified) to the analysis based on the {\color{black} ``per-event PDFs''},
which is the one yielding the best sensitivity.

\subsubsection{Sensitivity}
The sensitivity $\sens$ is calculated as the median of the distribution 
of the \SI{90}{\percent} C.L. upper limits computed for an ensemble of 
pseudo-experiments with a null-signal hypothesis (\fref{fig:sensitivity}). 
They are generated 
according to the PDFs constructed for RMD and ACC background and assuming the rates of  
the RMD and ACC events evaluated in the side-bands. 
The sensitivity is estimated to be 
$\sens = \num{8.8e-13}$.

The limits include 
{\color{black} systematic uncertainties 
with dominant contribution
from detector misalignment, 
\photon-ray energy scales
and normalisation.}

{\color{black} 
Misalignment between detectors is calibrated using tracks of particles from muon decays or cosmic rays, crossing multiple detectors. The target position is determined from a combination of muon decay point reconstruction and a photogrammetric method exploiting the cameras installed inside the magnet bore. The $\photon$-ray energy scale is calibrated with a combined analysis of CEX, CW, cosmic ray and side-band spectra. The worsening of the sensitivity due to the inclusion of systematic uncertainties is \SI{5.0(3.7)}{\percent}.}


\subsubsection{Event distributions and likelihood fit in the analysis region}

\begin{figure}[tbp]
\centering
  \includegraphics[width=20pc,angle=0] {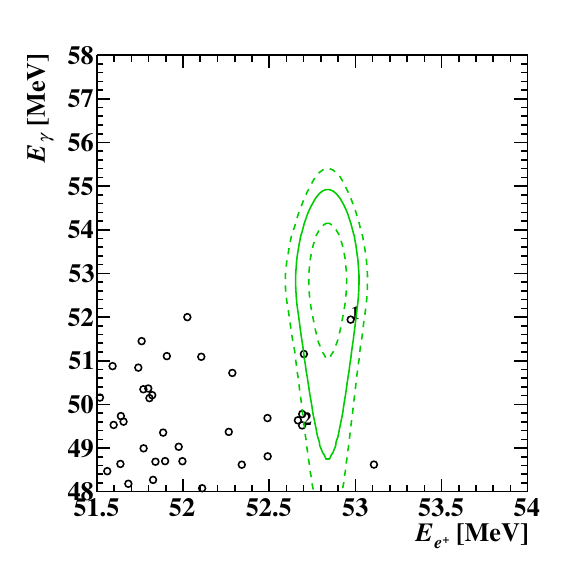}
  \includegraphics[width=20pc,angle=0] {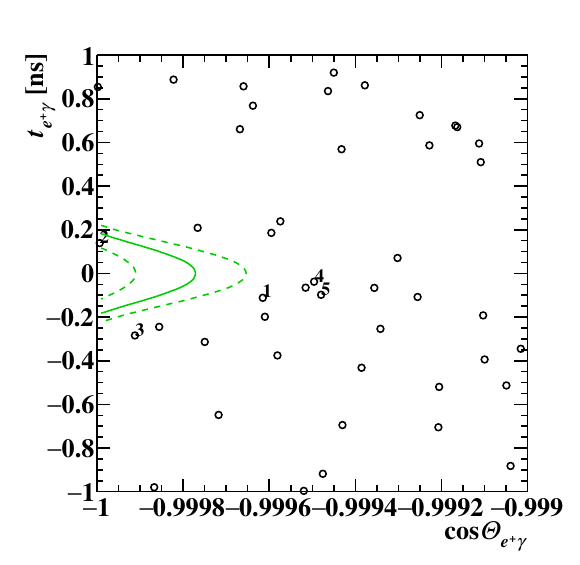}
 \caption{
Event distributions on the $(\epositron, \egamma)$- 
and $(\cos\Thetaegamma, \tegamma)$-planes. 
Selections of $\cos\Thetaegamma < -0.9995$ and 
$|\tegamma| < \SI{0.2}{\ns}$, which have \SI{97}{\percent} signal efficiency for each observable, 
are applied for the $(\epositron, \egamma)$-plane, 
while selections of $\num{49.0} < \egamma < \SI{55.0}{\MeV}$
and $52.5 < \epositron < \SI{53.2}{\MeV}$, 
which have signal efficiencies of \SI{93}{\percent} and \SI{97}{\percent}, respectively,
are applied for the $(\cos\Thetaegamma, \tegamma)$-plane.
The signal PDF contours ($1\sigma$, $1.64\sigma$ 
and $2\sigma$) are also shown.
The five highest-ranked events in terms of $\rsig$ are indicated 
in the event distributions, if they satisfies the selection.
}
\label{fig:distribution2D}
\end{figure}

A total of 66 events were observed in the analysis region.
The event distributions in the $(\epositron, \egamma)$ and $(\cos\Thetaegamma, \tegamma)$ planes are shown in~\fref{fig:distribution2D},
where even tighter selection requirements are applied to the discriminating variables to have a closer look around the signal region. The contours of the averaged signal PDFs are also shown for reference. No excess of events is observed in the region where the signal PDFs are peaking.

\begin{figure}[tbp]
\centering
\includegraphics[width=20pc] {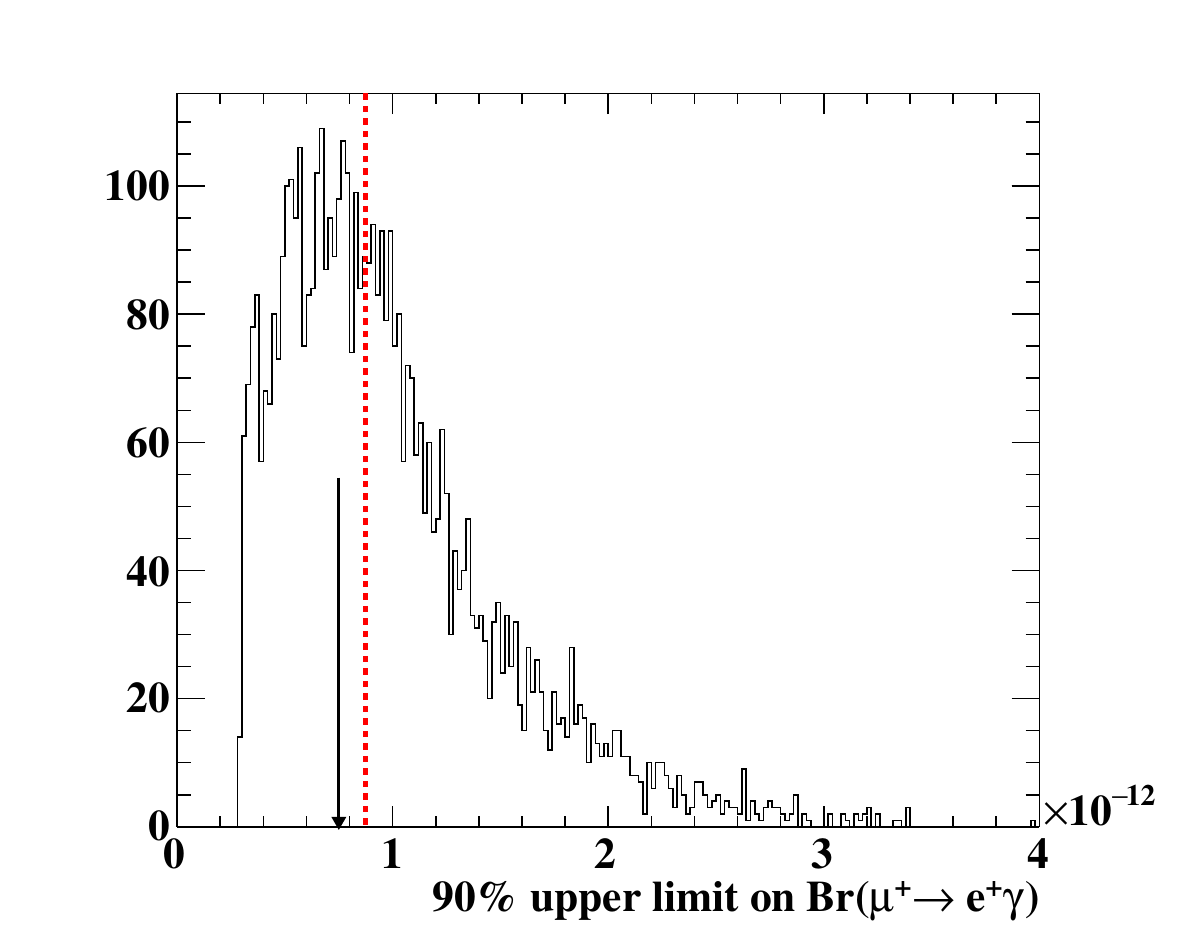}
\caption{\label{fig:sensitivity}
Distribution of the \SI{90}{\percent} C.L. upper limits computed for an ensemble of 
pseudo-experiments with a null-signal hypothesis. 
The sensitivity is calculated as the median of the distribution to be 
$\sens = \num{8.8e-13}$.
The sensitivity is indicated by a
red dashed line 
while the upper limit observed in the analysis region with a solid arrow.
}
\end{figure}

\begin{figure*}[htb]
\centering
\includegraphics[width=\textwidth,angle=0] {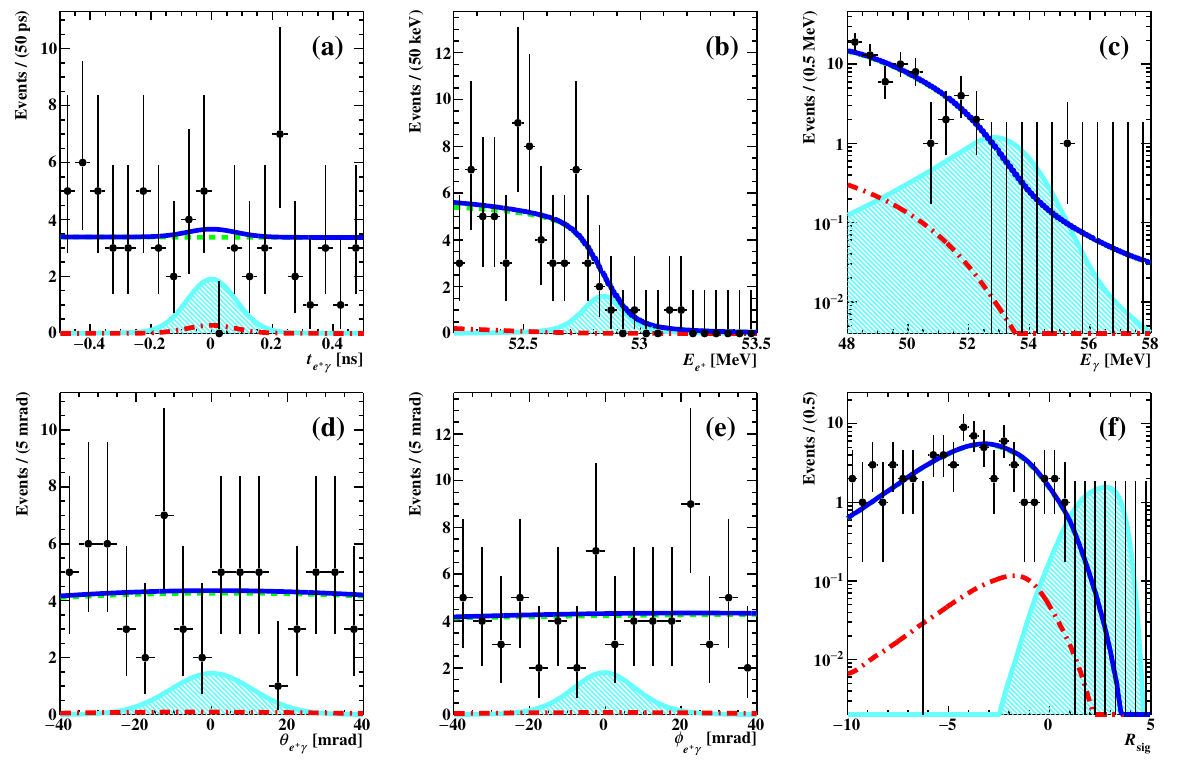}
 \caption{
The projections of the best-fitted PDFs to the five main observables and $\rsig$, together with the data distributions (black dots). 
The green dash and red dot-dash lines are individual components of 
the fitted PDFs of ACC and RMD, respectively. The blue solid line is the sum 
of the best-fitted PDFs. The cyan hatched histograms show the signal PDFs corresponding to four times magnified $\nsig$ upper limit.
}
 \label{fig:FitResult1D}
\end{figure*}

\Fref{fig:FitResult1D} shows the projected data distribution for each 
of the observables $(\epositron, \egamma, \tegamma, \thetaegamma, \phiegamma)$, for all events in the analysis region, with the best-fitted PDFs.
All data distributions are well-fitted by their 
background PDFs.
\Fref{fig:FitResult1D} (f) shows the data distribution of the relative signal likelihood $\rsig$, defined as 
\begin{equation*}
\rsig = \log_{10} \left( \frac{S(\vector{x}_i)}{f_\mathrm{RMD}R(\vector{x}_i)+f_\mathrm{ACC}A(\vector{x}_i)} \right) \; ,
\end{equation*}
where $f_\mathrm{RMD}$ and $f_\mathrm{ACC}$ are 
the expected fractions of the RMD and ACC background events, which 
are estimated to be \num{0.02} and \num{0.98} in the side-bands, respectively. 
The data distribution for $\rsig$ also shows a good agreement 
with the distribution expected from the likelihood fit result. 
The five highest-ranked events in terms of $\rsig$ are indicated 
in the event distributions shown in \fref{fig:distribution2D}. 

\begin{figure}[htb]
\centering
  \includegraphics[width=20pc,angle=0] {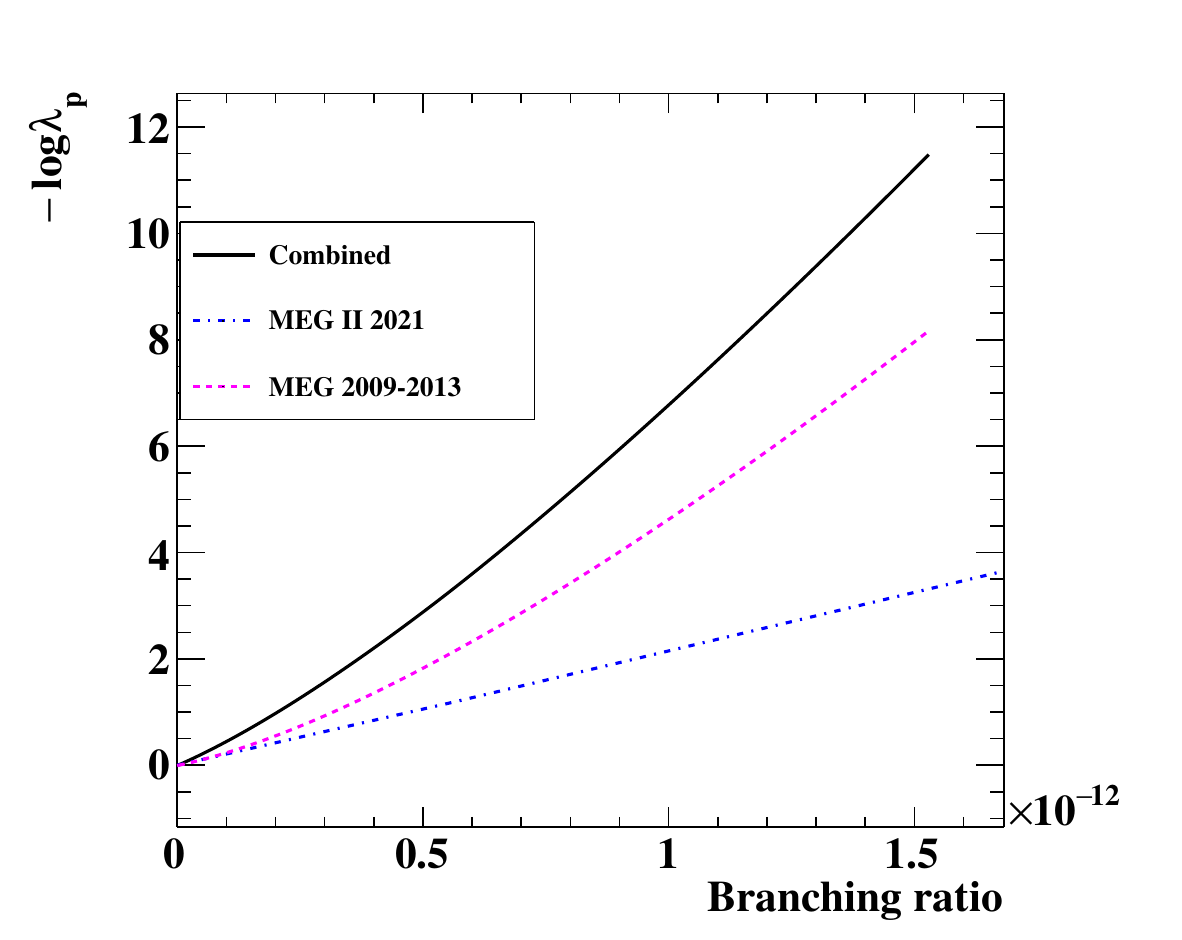}
 \caption{
 The negative log likelihood-ratio ($\lambda_{\rm p}$) as a function of 
    the branching ratio. 
    The three curves correspond to the MEG~II 2021 data, the MEG full dataset \cite{baldini_2016} and the combined result. 
}
 \label{fig:NLL}
\end{figure}

\Fref{fig:NLL} shows the observed profile likelihood ratio as a function of the branching ratio. 
The computation of the confidence interval with the Feldman--Cousins prescription, 
which is performed with the profile likelihood ratios for positive $\nsig$ only, 
is not affected by the behaviour of the curve at negative, nonphysical branching ratios. 
Nonetheless, for completeness, we also compute the likelihood ratio in the negative side, although we have to set the bound $\nsig > \num{-0.004}$ ($\BR > \num{-1.5e-15}$) (not distinguishable from zero in the figure) to ensure that the total PDF is always positive-valued all over the analysis region.
The best estimate and the \SI{90}{\percent} C.L. upper limit of the branching ratio
are estimated to be $\bestfit = \num{-1.1e-16}$ and $\ul = \num{7.5e-13}$, respectively.
The obtained upper limit is consistent with the sensitivity 
calculated from the pseudo-experiments with a null-signal hypothesis (\fref{fig:sensitivity}). 

The limit includes the systematic uncertainties, the impact of which is an increase by \SI{1.5}{\percent}, consistent within statistical uncertainties with what is expected from pseudo-experiments.

\subsubsection{Consistency checks}

\begin{figure}[htb]
\centering
  \includegraphics[width=20pc,angle=0] {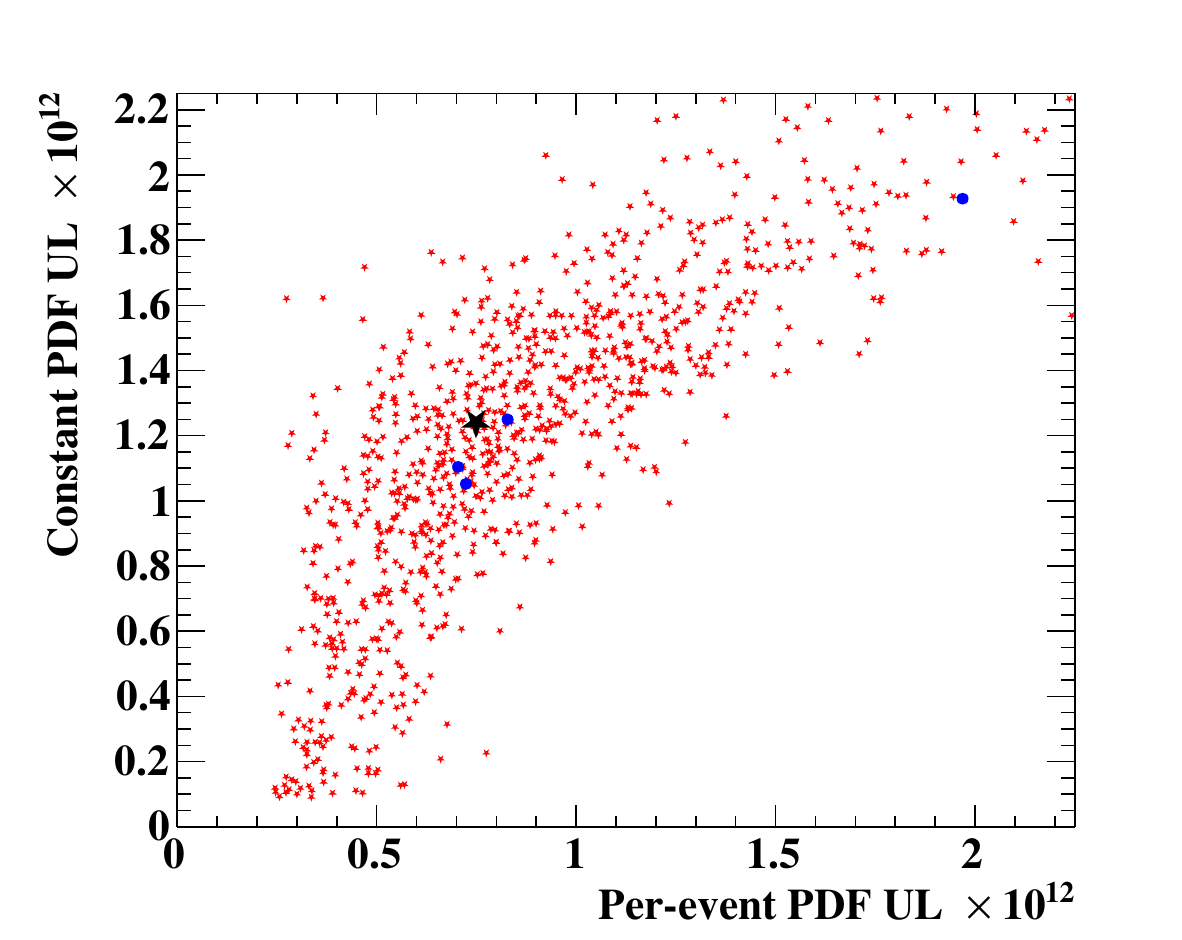}
 \caption{
 Comparison of the branching ratio upper limits (without systematic uncertainties) extracted by the two likelihood analyses when run over a common ensemble of pseudo-experiments (red dots). The results obtained on real data are also shown, for the analysis region (black star) and four fictitious analysis regions in the time side-bands: $\SI{-3}{\ns}<\tegamma<\SI{-2}{\ns}$, $\SI{-2}{\ns}<\tegamma<\SI{-1}{\ns}$, $\SI{1}{\ns}<\tegamma<\SI{2}{\ns}$, $\SI{2}{\ns}<\tegamma<\SI{3}{\ns}$ (blue dots).
}
 \label{fig:ULComparison}
\end{figure}

With the maximum likelihood analysis using the constant PDF approach, the \SI{90}{\percent} C.L. upper limit of the branching ratio, including systematic uncertainties, is $\ul = \num{1.31e-12}$.
The consistency between the results of the two analyses is checked on a common ensemble of pseudo-experiments generated with a null-signal hypothesis.
The comparison of the \SI{90}{\percent} C.L. upper limits obtained by the two analyses on the common pseudo-experiments is shown in \fref{fig:ULComparison}, where systematic uncertainties are not included for simplicity, resulting in slightly smaller upper limits.
The two results are strongly correlated, with the per-event PDFs' analysis showing \SI{\sim 30}{\percent} better sensitivity.
The upper limits obtained in the analysis region and in the fictitious analysis 
regions in the time side-bands 
are also shown in \fref{fig:ULComparison}, and are found to be in good agreement with the results of the pseudo-experiments.

To validate the techniques used to parameterise the signal PDFs, pseudo-experiments generated with a null-signal hypothesis were mixed with signal Monte Carlo samples coming from the Geant4 simulation of the full detector~\cite{megII-det}, assuming an expected signal yield of 10 events. A likelihood fit was performed, adopting the same techniques used on real   data to parameterise the PDFs, including the correlations. We obtained a distribution of best-fit values with an average consistent with $N_\mathrm{sig} = 10$, we checked the correct coverage of the confidence intervals, and verified the consistency of the $R_\mathrm{sig}$ distributions with the ones obtained exclusively from pseudo-experiments.

The analysis was also applied to four fictitious analysis regions inside the time side-bands ($-3<\tegamma<\SI{-2}{\ns}, -2<\tegamma<\SI{-1}{\ns}, 1<\tegamma<\SI{2}{\ns}, 2<\tegamma<\SI{3}{\ns}$). The results are also shown in~\fref{fig:ULComparison} and are consistent with the distribution of the \SI{90}{\percent} C.L. upper limits in the pseudo-experiments.

Finally, the likelihood fit in the analysis region was also performed without the constraints on 
$\nrd$ and $\nacc$. The best estimates of $\nrd=\num{0\pm 3.9}$ and $\nacc=\num{66\pm 8.1}$ are well consistent 
with the side-band estimates of $\num{1.2\pm 0.2}$ and $\num{68\pm 3.5}$, respectively.

\subsubsection{Combination with the MEG result}
The sensitivity of this analysis with the 2021 data is comparable to the one with the full MEG dataset \cite{baldini_2016}
despite the much smaller dataset thanks to improvements in resolutions and efficiencies.
The upper limit obtained in this analysis is combined with the MEG result.
The two results are combined in a simplified manner, setting a threshold on the negative log likelihood-ratio curve instead of 
following the Feldman--Cousins approach.
The negative log likelihood-ratio curves for the MEG full dataset and the MEG~II 2021 data, including systematic uncertainties, are shown in \fref{fig:NLL}, along with their combination,  coming from the product of the two likelihood functions. The upper limit is determined as the branching ratio value at which the combined curve crosses a threshold of \num{1.6}, which is conservatively chosen from pseudo-experiments, to match approximately the results of the Feldman--Cousins method. 
{\color{black} The systematic errors in MEG and MEG~II are small and weakly correlated. The effect of neglecting the correlation has a negligible effect compared to the approximation implicit in the approach to combine the results.}
The combined upper limit at \SI{90}{\percent} C.L. is computed to be $\ul = \num{3.1e-13}$.
It is consistent with the combined sensitivity of $\sens = \num{4.3e-13}$, 
which is estimated with the combined pseudo-experiments, 
with a \SI{30}{\percent} probability of having a more stringent upper limit.

\section{Conclusions and perspectives}

In 2021, the MEG~II experiment was commissioned and started taking data with \megp\ trigger for seven weeks.
A blind, maximum-likelihood analysis
found no significant event excess compared to the expected background and established a \SI{90}{\percent} C.L. upper limit on the branching ratio
${\cal B} (\megp) < \num{7.5e-13}$. 

When combined with the final result of MEG, we obtain the most stringent limit up to date, ${\cal B} (\megp) < \num{3.1e-13}$.

The MEG~II collaboration has continued to take data during 2022 and 2023, with a projected statistic ten-fold larger than in 2021,
and a more than twenty-fold increase in statistics is foreseen by 2026, with the goal of reaching a sensitivity to the $\megp$ decay of $\sens \num{\sim 6.0e-14}$.

\section*{Acknowledgments}

We are grateful for the support and cooperation provided  by PSI as the host laboratory and to the technical and engineering staff of our institutes.
This work is supported by 
DOE DEFG02-91ER40679 (USA); 
INFN (Italy); 
H2020 Marie Skłodowska-Curie ITN Grant Agreement 858199;
JSPS KAKENHI numbers JP26000004, 20H00154,21H04991,21H00065,22K21350 and JSPS Core-to-Core Program, A. Advanced Research Networks JPJSCCA20180004 (Japan);
Schweizerischer Nationalfonds (SNF) Grants 206021\_177038,
206021\_157742, 200020\_172706, 200020\_162654 and 200021\_137738 (Switzerland); the Leverhulme Trust, LIP-2021-01 (UK).
\bibliographystyle{my}
\bibliography{MEGIIPhysicsPaper2021}
\end{document}